\documentclass[a4paper,11pt]{article}
\usepackage{pos}
\usepackage{epsfig}
\newcommand{\postscript}[2]{\setlength{\epsfxsize}{#2\hsize}
   \centerline{\epsfbox{#1}}}

\title{High-Energy Neutrinos from NGC 1068}

\ShortTitle{High-Energy Neutrinos from NGC 1068}

\author*[a,b,c]{Luis A. Anchordoqui}
\author[d,e]{John F. Krizmanic}
\author[d,f]{Floyd W. Stecker}

\affiliation[a]{Department of Physics, Lehman College, City University of New
  York, NY 10468, USA}
\affiliation[b]{Department of Physics, Graduate Center, City University
  of New York,  NY 10016, USA}
\affiliation[c]{Department of Astrophysics, American Museum of Natural History, NY
10024, USA}
\affiliation[d]{NASA/Goddard Space Flight Center,
Greenbelt, MD 20771, USA}
\affiliation[e]{Department of Physics, University of Maryland, Baltimore, MD
21250, USA}
\affiliation[f]{Department of Physics and Astronomy,
University of California, Los Angeles, CA 90095, USA}

\abstract{\noindent IceCube has observed an excess of neutrino
  events over expectations from the isotropic background from the
  direction of NGC 1068. The excess is inconsistent with background
  expectations at the level of $2.9\sigma$ after accounting for
  statistical trials. Even though the excess is not statistical significant yet, it is interesting to entertain the possibility that it corresponds to a real signal. Assuming a single power-law spectrum, the
  IceCube Collaboration has reported a best-fit flux $\phi_\nu\sim 3 \times
  10^{-8}  (E_\nu/{\rm TeV})^{-3.2}~({\rm GeV \, cm^2 \, s})^{-1}$, where $E_\nu$ is the neutrino energy. Taking account of new physics and astronomy developments we give a revised high-energy neutrino flux for the Stecker-Done-Salamon-Sommers AGN core model and show that it can accommodate IceCube observations.}

\FullConference{37$^{\rm{th}}$ International Cosmic Ray Conference (ICRC 2021)\\
		July 12th -- 23rd, 2021\\
		Online -- Berlin, Germany}

\begin{document}
\maketitle

\begin{figure}
        \postscript{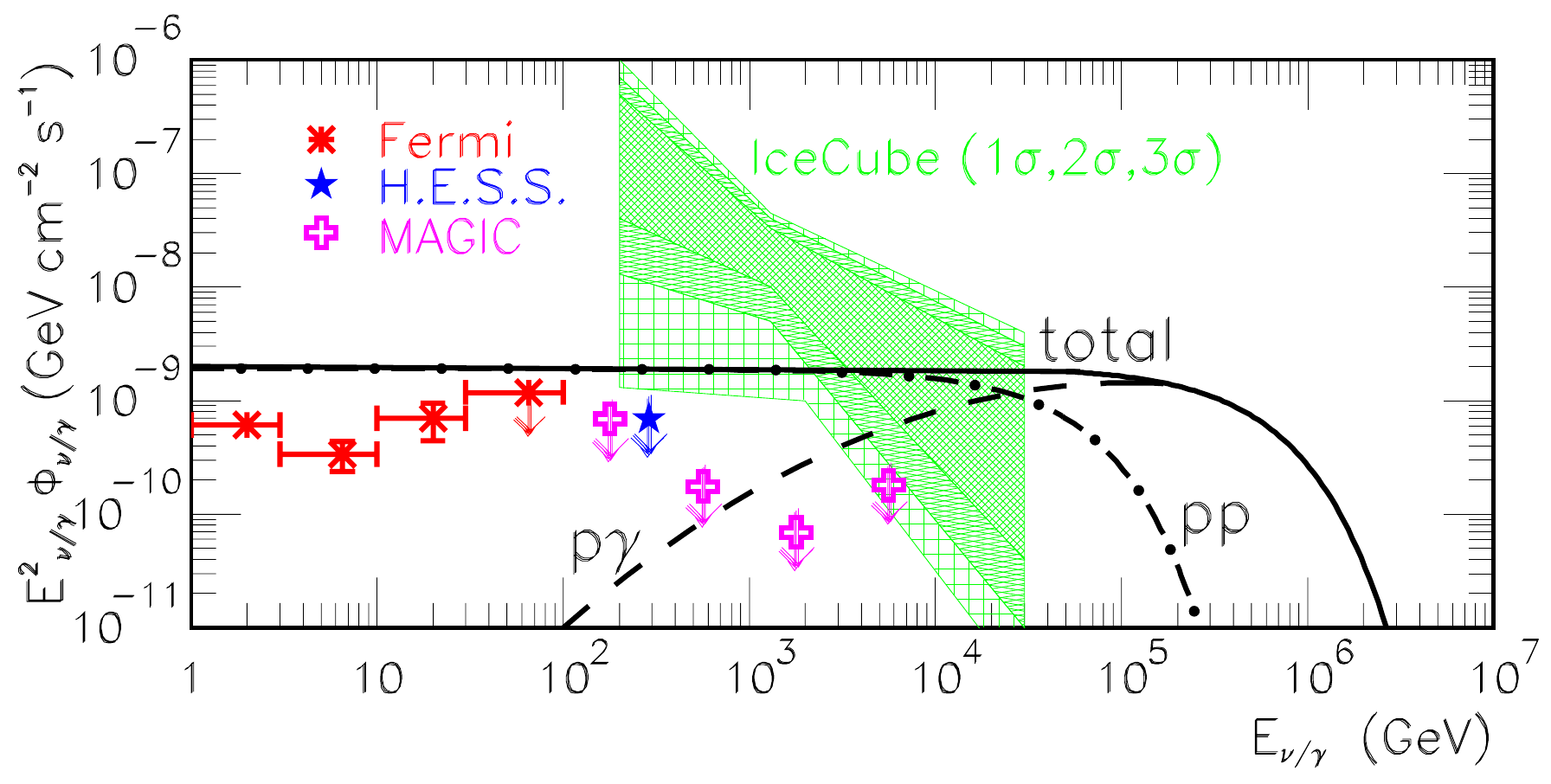}{0.8}
        \caption{AGN-core model prediction of the muon neutrino
          spectrum from NGC 1068; $pp$ interactions (dot-dashed line),
          $p\gamma$ interactions (dashed line), and total (solid
          line). For comparison, we overplot the best-fit
          time-integrated astrophysical power-law neutrino flux
          obtained using the 10~yr IceCube
          data~\cite{Aartsen:2019fau}.  We also show measurements and
          bounds on the gamma-ray flux from
          Fermi~\cite{Lamastra:2016axo},
          H.E.S.S.~\cite{Aharonian:2005ar} and
          MAGIC~\cite{Acciari:2019raw}. We have set $b=1$ in our
          calculations. \label{fig:1}}
      \end{figure}

A search for astrophysical point-like neutrino sources using 10~yr of data collected by the IceCube detector (between April 6, 2006 and
July 10, 2018) finds an excess of clustered events (with energies
$E_\nu \gtrsim 1~{\rm TeV}$) over expectations from an isotropic sky coincident with
the Seyfert II galaxy NGC 1068~\cite{Aartsen:2019fau}.  The excess is inconsistent with
background expectations at the level of $2.9\sigma$ after accounting
for statistical trials. When the distributions of 
  the observed events as a
  function of their distance from NGC 1068 and
  their estimated angular uncertainties
are weighted by a signal-over-background likelihood characterizing the point-like source hypothesis give a best fit spectrum $\propto
E_\nu^{-3.2}$. On the assumption of a single power-law spectrum
  IceCube finds a best-fit flux $\phi_\nu \sim 3 \times
  10^{-8}  (E_\nu/{\rm TeV})^{-3.2}~({\rm GeV \, cm^2 \, s})^{-1}$; the reconstructed muon neutrino spectrum with its large
uncertainty is shown in Fig.~\ref{fig:1}. A point worth noting at this
juncture is that the favored soft  spectrum for NGC 1068  is consistent
with the shape of the high-energy starting event  all-sky neutrino
spectrum, which is
compatible with an unbroken power-law spectrum, with a preferred
spectral index of $2.87^{+0.20}_{-0.19}$ for the 68.3~\% confidence interval~\cite{Abbasi:2020jmh}.

Recently, the MAGIC Collaboration reported a search for gamma-ray
emission in the very-high-energy band~\cite{Acciari:2019raw}. No
significant signal was detected during 125 hours of observation of NGC
1068. The null result provides a 95\% CL upper limit to the gamma-ray
flux above $200~{\rm GeV}$ of
$5.1 \times 10^{-13}~({\rm cm^2 \, s})^{-1}$. This limit improves an
earlier upper bound from H.E.S.S.~\cite{Aharonian:2005ar} and set tight constraints
on the theoretical models that could explain the NGC 1068 IceCube's
``signal.'' More concretely, the gamma rays accompanying the neutrino
flux must be significantly attenuated; see Fig.~\ref{fig:1}.  The gamma-ray optical depth is
well-known,
\begin{equation}
  \tau_{\gamma\gamma} (\varepsilon) \sim \frac{\sigma_{\gamma \gamma}}{4 \pi c}
    \frac{L_X}{\varepsilon R} \sim 10^5
   \ \left(\frac{\varepsilon}{1~{\rm keV}}\right)^{-1}  \ \left(\frac{L_X}{L_{\rm
        Edd}}\right) \ \left(\frac{R_S}{R} \right) \,,
\label{taugg}
\end{equation}      
where $\varepsilon$ is the typical energy of the target photon
background, $\sigma_{\gamma \gamma}$ is the scattering cross section,
$L_X$ is the $X$-ray luminosity, $R$ is the size of the region carrying the dense $X$-ray target photons, $R_S = 2 GM/c^2$ is the is the Schwarzschild radius,  $L_{\rm Edd} = 4 \pi GM m_p
      c/\sigma_T$ is the Eddington luminosity (i.e., the maximum steady state luminosity that
      can be produced before radiation pressure disrupts the accretion
      flow), and $\sigma_T$ is the Thomson cross section, with $M$ and $m_p$ the
      black hole and proton mass, respectively. It is straightforward
      to see using  (\ref{taugg}) that to have significant absorption
      of the gamma rays $R$ must characterize a compact region in the vicinity of the black hole.

In 1991, Stecker, Done, Salamon, and Sommers proposed a model
featuring all of these characteristics~\cite{Stecker:1991vm}. The high-energy neutrino flux
originates in the core of the active galactic nucleus (AGN). Protons can reach very high energies through accretion-disk
shock-acceleration at the inner edge of the black hole~\cite{Protheroe}.  The
relativistic protons undergo inelastic collisions with the
thermal photon background to produce charged and neutral pions,
which, in turn, decay into neutrinos, electrons, and gamma-rays. The 
non-thermal electrons generate gamma rays via inverse Compton
scattering on disk photons. While all the gamma rays cascade down to the MeV
energy range because of the strong internal attenuation
effect~\cite{Inoue:2019fil} neutrinos escape the source {\it en
  route} to Earth. Taking account of new physics and astronomy developments, in this
communication we give a revised high energy neutrino flux for the AGN
core model and show that it can accommodate IceCube data.  Observational studies and theoretical modeling are used to guide us in
choosing the model parameters.  Before proceeding, we pause to note that related ideas for modelling neutrino emission from NGC 1068 have been discussed in~\cite{Murase:2019vdl,Inoue:2019yfs,Muller:2020hvs,Kheirandish:2021wkm}.      

$X$-ray absorbers are classified as Compton-thick or -thin, according to
whether their column density $N_{\rm H}$ is larger or smaller than $\sigma^{-1}_T \simeq
1.5 \times 10^{24}~{\rm cm}^{-2}$. Back in the 90s, there was a lack of evidence for strong $X$-ray absorption features in AGN
spectra~\cite{Mushotzky,Turner:1980mq}, and this was taken as an indication that the secondary $X$-rays are
produced in regions of low column density. If this were the case, the amount of
target gas for $pp$ collisions would be very limited and the very large
photon density in the AGN core would make photopion production, predominantly through the resonant process
$p \gamma \to \Delta^+ \to n \pi^+$ or $p \pi^0$, the leading mechanism for energy loss. Because of resonant scattering the mean pion energy is kinematically determined by requiring equal boosts for
the decay products of the $\Delta^+$, giving $\langle E_\pi\rangle
\sim E_p/5$~\cite{Stecker:1968uc}. Likewise, to a first approximation we relate the energy of the neutrinos with that of the
parent protons considering that the four (massless) particles
resulting from the decay $\pi^+ \to \mu^+ \nu_\mu \to e^+ \nu_e  \overline
\nu_\mu \nu_\mu$ (and the charge-conjugate processes) share similar amounts of energy $\langle E_\nu \rangle \simeq
\langle E_\pi\rangle/4 \simeq
E_p/20$. For the neutral pions $\pi^0  \to \gamma
\gamma$,  we similarly find $\langle E_\gamma \rangle \simeq E_p/10$.
The threshold condition for pion
production in $p\gamma$ scattering is given by $(p_\gamma + p_p)^2 > (m_p +
m_\pi)^2$, which leads to 
$\zeta >  (2 m_\pi m_p +m_\pi^2)/m_p^2 \equiv
\zeta_0 \simeq 0.313$, where the dimensionless variable $\zeta \equiv
4 \varepsilon E_p/m_p^2 $ characterizes the center-of-mass total
energy squared of the interaction and where we have taken $m_\pi^\pm \simeq m_\pi^0
\simeq 137~{\rm MeV}$ and $m_p \simeq m_n \simeq 938~{\rm MeV}$. For UV photons, with a mean energy  $\langle \varepsilon  \rangle
\sim 40~{\rm eV}$, this translates into a characteristic proton energy
$E_{p,{\rm min}} >  70~{\rm PeV}/(\varepsilon/{\rm eV}) \sim 2~{\rm PeV}$~\cite{Stecker:1991vm}. The fact that this reaction turns on at so high energies implies that
the photons and neutrinos from decaying pions are produced at very
high energies too, well above the TeV range. Thereby, IceCube's
observation of ${\cal O} ({\rm TeV})$ neutrinos from the direction of NGC 1068 pose unique challenges for predictive modeling.

Over the past decades, multiple space-missions and ground-based
experiments (including BeppoSAX, Chandra, MERLIN, the Very Long
Baseline Array, NuSTAR, and
XMM-Newton~\cite{Guainazzi:1999ct,Young:2001ck,Smith:2003dc,Gallimore:2004wk,Bauer:2014rla,Marinucci:2015fqo})
have performed an extensive observing campaign aimed at the
characterization of NGC 1068. Collectively,
these observations call for a recalibration of the AGN-core-model parameters. In
particular, NuSTAR detected a flux excess above 20~keV with respect to
both the December 2012 observation and a later observation performed
in February 2015. The most plausible explanation of the NuSTAR
transient excess is that for a short time interval the total absorbing
column, probably composed by a number of individual clouds, became
less thick  so as to allow  the radiation from the AGN core to pierce
through it, supporting the hypothesis of a clumpy structure of the
obscuring material along the line of sight. The inferred column gas
density from NuSTAR observations, which varies in the range $5.9
\times 10^{24} \lesssim N_{\rm H}/{\rm
  cm^2} \lesssim  8.5 \times 10^{24}$~\cite{Marinucci:2015fqo}, seems to indicate that
the target proton gas in the AGN core is much denser than previously thought~\cite{Turner:1980mq}. If this
were the case, NGC 1068 should  
be reclassified as an optically thick absorber. For $pp$ collisions, threshold effects are
insignificant and so  for column densities $N_H > \sigma_T^{-1}$, $pp$ scattering could produce a TeV neutrino
population to explain the low-energy tail of IceCube's
``signal''~\cite{Aartsen:2019fau}. Moreover, after  correction for absorption, the inferred intrinsic $X$-ray luminosity of NGC 1068 (in
the $2 - 10~{\rm keV}$ range) is
$L_X = 6^{+7}_{-4} \times 10^{43}~{\rm erg \
  s^{-1}}$~\cite{Marinucci:2015fqo}, above about  2 orders of
magnitude than previous estimates~\cite{Young:2001ck}. It is important to note that the intrinsic $L_X$ of NGC 1068 is roughly an order of magnitude larger than the
$L_X$ of NGC 4151~\cite{Piccinotti,Singh:2011nx}, which
is the brightest Seyfert in $X$-rays. Since both these two sources are located
at about 14~Mpc from Earth~\cite{Tully}, the $L_X$ recalibration of~\cite{Marinucci:2015fqo} makes NGC 1068 the intrinsically brightest Seyfert galaxy in the sky, and explains why it
could become the first neutrino source to be uncovered using (only)
IceCube data.\footnote{Multimessenger observations of TXS 0506+056
  provided $3\sigma$ evidence
  of neutrino emission from the flaring
  blazar~\cite{IceCube:2018dnn}. However, the association of the Texas
  source with neutrino emission in IceCube's 10~yr data sample is less significant~\cite{Aartsen:2019fau} than the reported significance of the time-dependent flare
associating both neutrino and gamma-ray production.}

To develop some sense for the orders of magnitude involved, we begin by noting that first-order Fermi acceleration of protons
in strong (non-relativistic) shocks produces a power-law proton energy
spectrum $\propto E_p^{-2}$ up to a maximum energy $E_{p, {\rm
    max}}$.  The proton acceleration time-scale is given by
  \begin{equation}
    t_{\rm acc} (E_p) \sim 5 \times 10^{-2} \ b \ \left(\frac{R_{\rm
          shock}}{R_S}\right) \left(\frac{B}{{\rm G}}\right)^{-1}
      \left(\frac{E_p}{m_p} \right)~{\rm s}\,, 
\label{eq:acc}
    \end{equation}
      where $B \simeq 5.5
      \times 10^{27} Q^{-1/2} (R_{\rm shock}/R_S)^{-7/4} \,
      L_X^{-1/2}~{\rm G}$ is the magnetic field, $R_{\rm shock}$ is the shock
      radius, $Q = 1- 0.1 (R_{\rm shock}/R_S)^{0.31}$ is the
      efficiency of conversion of bulk kinetic energy of accreting
      plasma into energetic particles at the shock, and $b$ is a numerical
      factor that gives a measure of the particle's mean free path (in
      gyroradii) for scattering off the magnetic field inhomogeneities~\cite{Ellison:1983zzd,Szabo:1994qx}. Based on the assumption $L_X \sim
      L_{\rm Edd}/20$ (which corresponds to $M \sim 10^7 M_\odot$)  we fix the shock radius to
      $R_{\rm shock} \sim 10 R_S$~\cite{Stecker:1991vm}. 

The $pp$
      energy-loss rate is given by
      \begin{equation}
        t_{pp} (E_p) = \frac{1}{n_p \ \sigma_{pp} c \kappa_{pp}}  \,,
\label{eq:losspp}
      \end{equation}
      where $\sigma_{pp} (E_p) \sim [34.3 + 1.88 \ln(E_p/{\rm TeV}) +
      0.25 \ln^2(E_p/{\rm TeV})] \times 10^{-26}~{\rm
  cm}^2$ is the inelastic $pp$ cross section~\cite{Kelner:2006tc}, $\kappa_{pp} \sim 0.5$ is the proton inelasticity of the
      process~\cite{Frichter:1997wh}, and $n_p \sim N_{\rm H}/R$ the
      mean proton density. Following~\cite{Stecker:1991vm}, we take $R \sim 30 R_S$.

The $p\gamma$ energy-loss rate is evaluated by
\begin{equation}
 t_{p\gamma}^{-1} (E_p) = \frac{c}{2} \,\int_0^\infty
                 d\varepsilon \,\frac{n(\varepsilon)}{\gamma^2 \varepsilon^2}\, \int_0^{2\gamma\varepsilon}
                 d\varepsilon^\prime \, \varepsilon^\prime\,
                 \kappa_{p\gamma} \, \sigma_{p\gamma}(\varepsilon^\prime),
\label{eq:interaction}
\end{equation}
where $\gamma = E_p/m_pc^2$ is the Lorentz  boost, $\varepsilon'$ is
the photon energy in the proton rest frame, $n(\varepsilon)$ is the
differential number density of photons, and $\sigma_{p\gamma}$ and
$\kappa_{p\gamma}$ are the cross section and inelasticity for
photopion production, respectively~\cite{Stecker:1969fw}. We approximate the $p\gamma$ cross
section by interactions with the $\Delta^+$ resonance of mass
$m_\Delta \simeq 1.232~{\rm GeV}$. Since the decay width $\Gamma_\Delta \simeq
150~{\rm MeV}$ is much smaller than the resonance mass the cross section can be safely
approximated by the single pole of the narrow-width approximation,
\begin{equation}
\sigma_{p\gamma} (\varepsilon') = \pi \ \sigma_0\,\,
\frac{\Gamma}{2} \ 
\delta(\varepsilon' - \varepsilon_0)\, ,
\label{sigma}
\end{equation}
where $\sigma_0 \simeq 5 \times 10^{-28}~{\rm cm}^2$ is the resonance peak
and $\varepsilon_0 = (m_\Delta^2 - m_p^2)/(2m_p) \simeq 340~{\rm MeV}$ the pole.  The factor of $\pi/2$ is
introduced to match the integral
(i.e. total cross section) of the Breit-Wigner and the delta
function. The photopion cooling rate  can now be readily
obtained  substituting (\ref{sigma}) into (\ref{eq:interaction}),
\begin{eqnarray}
  t_{p\gamma}^{-1} (E_p) & \approx & \frac{c\, \pi\,
    \sigma_0
    \,\varepsilon_0\, 
\Gamma_\Delta \, \kappa_{p\gamma}}{4\,
    \gamma^2}
  \int_0^\infty \frac{d \varepsilon}{\varepsilon^2}\,\,\, n(\varepsilon) \,\,\,
                                     \Theta (2 \gamma \varepsilon - \varepsilon_0)
=\frac{c \, \pi \, \sigma_0 \,\varepsilon_0\, \Gamma_\Delta \, \kappa_{p\gamma}}{4 \gamma^2}
  \int_{\epsilon_0/2 \gamma}^\infty \frac{d\varepsilon}{\varepsilon^2}\,\,
  n (\varepsilon) \nonumber \\
  & = & 
  \frac{c\, \pi \ \sigma_0 \ (m_\Delta^2 - m_p^2 ) \ \Gamma_\Delta \, \kappa_{p\gamma}}{
    8 \ m_p }  \ \left(\frac{m_p}{E_p}\right)^2
  \int_{\varepsilon_{\rm min}}^\infty \frac{d \varepsilon}{\varepsilon^2} \, n_\gamma
  (\varepsilon),
\label{eq:losspg}
\end{eqnarray}
where $\varepsilon_{\rm min} = (m_\Delta^2 -
m_p^2)/(4E_p)$~\cite{Anchordoqui:2018qom}. We assume that the spectrum
of the external UV radiation field arises from a Shakura-Sunyaev
optically-thick accretion disk model that is scattered by
clouds~\cite{Shakura:1972te}. For calculations, we approximate the AGN
continuum 
$n(\varepsilon)$ by two components: {\it (i)}~a power-law spectrum $\propto \varepsilon^{-1.7}$
which extends up to 1~MeV  and {\it (ii)}~a black body spectrum with
temperature $T = 5 \times 10^4~{\rm K}$ used to represent the
UV/optical bump which is thought to be thermal emission from the
accretion disk~\cite{Szabo:1994qx}. For normalization, we assume that
the total $X$-ray luminosity is roughly the same as that in the
UV-bump  $L_X \sim L_{UV}$ and so $L_C \sim 4 \pi R^2 c \int
\varepsilon n(\varepsilon) d \varepsilon = L_{\rm
  Edd}/10$, where $L_C$ is the luminosity in the infrared to hard
$X$-ray continuum~\cite{Stecker:1991vm}.

Now,  by equating 
(\ref{eq:acc}) to  (\ref{eq:losspp})+ (\ref{eq:losspg})  with $b=1$ it
is easily seen that $E_{p,{\rm max}}$ is ${\cal O}  (10^7~{\rm
  GeV})$. The order of magnitude estimate from this
back-of-the-envelope calculation is consistent with the result from a Monte Carlo simulation, which gives
\begin{equation}
E_{p,{\rm max}} \simeq 1.8 \times 10^{7}~{\rm GeV} \, (6/b^2)^\alpha \,,
\label{Epmax}
\end{equation}
where $\alpha = 0.52$ for $b^2 <6$ and $\alpha = 0.18$ for $b^2
>6$~\cite{Protheroe:1992qs}. Armed with (\ref{eq:acc}),
(\ref{eq:losspp}), (\ref{eq:losspg}), and (\ref{Epmax}), together with
the inclusive pion spectra and the energy spectra of photons and leptons produced at $pp$~\cite{Kelner:2006tc,Stecker:1978ah} and
$p\gamma$~\cite{Stecker:1978ah,Kelner:2008ke} collisions it is straightforward to
calculate the muon neutrino yield from NGC 1068. Our results are
encapsulated in Fig.~\ref{fig:1}. At low energies the spectrum
$\propto E_\nu^{-2}$ from $pp$ interactions dominates; at high
energies the spectrum from $p\gamma$ interactions dominates. Corrections due to kaon decay and
threshold effects are
${\cal O} (10\%)$~\cite{Roulet:2020yye} and fall within erros. We have accounted for a
reduction in the muon-neutrino flux at production by a factor of 2 due
to neutrino
oscillations (whose discovery was made after the publication
of~\cite{Stecker:1991vm}). From (\ref{taugg}) we can see that the accompanying
photons from $\pi^0$ decay cascade down to lower energies, in
agreement with the upper limits from H.E.S.S.~\cite{Aharonian:2005ar}
and MAGIC~\cite{Acciari:2019raw}.

We now turn to compare our results with recent estimates of the
neutrino flux from NGC 1068. The predicted neutrino flux is in
agreement with the estimates of~\cite{Murase:2019vdl,Inoue:2019yfs,Kheirandish:2021wkm}.   However, it is important to stress
that the acceleration rate adopted in our study is significantly faster than the one
used in~\cite{Murase:2019vdl,Inoue:2019yfs,Kheirandish:2021wkm}. This
implies that the maximum energy is always controlled by $p\gamma$
interactions. In particular, $t_{p \gamma} (E_{p,{\rm max}}) \ll t_{pp} (E_{p,{\rm
    max}})$ even when considering the upper bound of $n_p \sim 9
\times 10^{10}~{\rm cm}^{-3}$. For the acceleration mechanisms entertain in~\cite{Murase:2019vdl,Inoue:2019yfs,Kheirandish:2021wkm}, the
column density cannot (significantly) surpass $\sigma_T^{-1}$
otherwise $pp$ collisions would control and largely reduce
$E_{p,{\rm max}}$. The neutrino flux predicted by the AGN-core model
is about an order of magnitude larger
than the estimate in~\cite{Muller:2020hvs}, which is normalized to
accommodate gamma-ray observations.

Although there are a few other nearby  AGN of this
magnitude which can potentially be detected as point sources, one
can integrate over the estimated AGN population out to the
horizon to obtain a prediction for the diffuse neutrino flux. The
result is simple: $\Phi_\nu \sim \frac{1}{4\pi} \,\,{\cal R}\,
  \,n_{_{\rm AGN}} \, \langle L_{\nu} \rangle,$ where ${\cal R} \simeq$~1 horizon $\simeq 3$~Gpc,
$n_{_{\rm AGN}} \sim 800~{\rm Gpc}^{-3}$ is the number density of AGN
with $L_X > 10^{43}~{\rm erg/s}$~\cite{Urry:1991wcicrc}, and
$\langle L_\nu \rangle$ is an average AGN neutrino luminosity (all
flavors). What has become of the energy red-shifting of the neutrino? A more careful calculation must include an additional factor,
$H_0 \int dz \ H^{-1}(z) \ L_\nu(z)/L_\nu(0)$, to account for effects
of the expanding universe ({\it viz.}, loss of energy associated with the
redshift $z$ and also depending on a choice of Hubble parameter $H$)
and possible source
evolution~\cite{Stecker:1991vm,Stecker:2005hn,Stecker:2013fxa}. However,
given the large uncertainty in the energy spectrum, we will ignore this order
of magnitude ``correction'' and just note that if $\langle L_{\nu}
\rangle \sim 10^{-2} \, L_X \, E_\nu^{-2}$, the diffuse
neutrino flux expected on Earth from the AGN population, $ E_\nu^2 \
\Phi_\nu \sim 10^{-8}~{\rm GeV \, (cm^2 \, sr \, s})^{-1}$,
would be in the ballpark of IceCube observations~\cite{Abbasi:2020jmh}. Curiously though,
there is a seemingly bumpy-structure in the spectrum of the
high-energy starting event sample around
the 
100~TeV energy bin. Coincidentally, this is the energy range in which
photopion production on the disk photons turns on.  It is then
tempting to speculate that if not all AGN are Compton-thick we
would expect a bump in the spectrum when AGN sources producing
neutrinos only via $p\gamma$ interactions come into play.

In summary, IceCube has detected an intriguing excess of events above
the isotropic background from the direction of NGC 1068. We have shown
that the origin of
these neutrinos can be traced back to a Fermi engine at the core of
this AGN. Absorption and interactions intrinsic to the source due to
the high opacity,  will result in a suppressed TeV gamma-ray flux to
accommodate H.E.S.S. and MAGIC upper limits. The neutrino AGN-core model is fully predictive  and will be confronted with
future IceCube data.

\section*{Acknowledgments}
LAA is supported by NSF
    Grant PHY-1620661 and NASA Grant 80NSSC18K0464. JFK is supported
by NASA Grant 80NSSC19K0626. FWS is suppoorted by NASA Fermi Grant 80NSSSC20K0413.

\end{document}